\documentclass[twocolumn,prd,aps,amsmath,amssymb,epsfig]{revtex4}
\usepackage{graphicx}
\usepackage{epsfig,rotate,latexsym}
\usepackage{hyperref}
\usepackage[arrowdel]{physics}

\def \be{\begin{equation}}
\def \ee{\end{equation}}
\def \nn{\nonumber}

\begin{document}
\title{Probing shapes of microbes using liquid crystal textures}

\author{Ajit M. Srivastava}
\email{ajit@iopb.res.in}
\affiliation{Institute of Physics, Sachivalaya Marg, 
Bhubaneswar 751005, India, \\
Homi Bhabha National Institute,
Training School Complex, Anushakti Nagar, Mumbai 400085, India}
%

\begin{abstract}
We propose a novel technique to probe shape of a single microbe embedded
in a nematic liquid
crystal (NLC) sample by observing geometry of dark brushes with optical
microscope using a cross-polarizer set up.  Assuming certain anchoring
conditions for the NLC director at the surface of the microbe, we determine
the resulting shapes of brushes using numerical simulations. Our results
suggest that for asymmetrical microbes (such as cylindrical shaped
bacteria/viruses), resulting brushes may carry the imprints of this
asymmetry (e.g. the aspect ratio of cylindrical shape) at relatively large
distances to be able to be seen using simple optical microscopy even for
microbe sizes  in few tens to few hundred nanometer range.
\end{abstract}

\maketitle

\section{Introduction}

Topological defects arise in a variety of physical systems. Vortices in 
superfluid helium, flux tubes in superconductors, monopoles and strings
in liquid crystals are some examples of topological defects. A remarkable
property of these objects originates from the word topology itself which
refers to global properties of the relevant system. Topological defects
in a system have {\it global} signature, even if the actual core of the
defect may be microscopic. The stability of these objects stems from
some topological invariant, such as the winding number of vortices or 
monopoles which refers to winding of some order parameter field in the
order parameter space as one traces a closed curve (or surface, for monopoles)
in the physical space. Topological defect present at some point leads to 
non-zero value of this winding number along a closed loop (or surface) in
the physical space, irrespective of the size of such a loop (surface).
Thus, in principle, presence of topological defect can be detected by
making measurements very far from the defect, never ever going close
to the defect core itself. 
However, for many cases, measurement of this winding number, by actually
tracing the winding of the order parameter field is not possible. This is
the situation for superfluid helium and superconductors where the relevant
degree of freedom of the order parameter field relates to the phase of
some quantum  wave function which is not directly detectable.  

 The situation is very different for topological defects in liquid crystals.
here the winding refers to actual rotation of local ordering of (say, rod like)
molecules which can be directly determined using optical microscope and a 
cross-polarizer set-up due to birefringence of liquid crystalline phase
\cite{books}. 
Observations of liquid crystal defects with such a system is routinely 
carried out and typically topological defects are identified in terms
of dark brushes originating from the location of defect core. A non-zero 
winding of the order parameter (the director, which refers to the local 
orientation of liquid crystal ordering) around the core of the defect 
extends to all regions enclosing the defect, resulting in brushes extending to
the edge of the entire system (for a single isolated defect), allowing for
easy observation using optical microscope. For typical liquid crystal defects,
the core size can be of order tens of nanometers, which obviously cannot be 
seen with an optical microscope. But the dark brushes emanating from such
a defect, being macroscopic, are easily seen, and geometry/topology of
these brushes allows for identification of the defect core (including the
sign of defect using rotation of the sample under microscope and observing
the sense of rotation of brushes \cite{books}).

 The crucial element of this topological information is the orientation of
the {\it director} at the core. Thus, for example for monopoles (hedgehogs)
in nematic liquid crystals (NLC), the director points radially outward at the 
surface of a spherical core. Thus, the director at the surface of the core 
traces a non-trivial winding on the order parameter space (which is RP$^2$
for nematics). This non-trivial winding, being topological, extends to all 
closed surfaces enclosing the hedgehog, and is identified easily in terms of
four dark brushes originating from a point like region when viewed with 
cross-polarizer set-up in an optical microscope. It is then obvious that
exactly the same structure of brushes will result if there was a microscopic
spherical object (say of similar size as the core of hedgehog, or bigger)
on whose surface the director assumes normal boundary condition. Liquid crystal
order parameter, the director, typically anchors at some fixed angles on 
surfaces of external objects (or, even medium such as NLC-air interface). 
Many surfaces allow normal boundary conditions. Though, with little thought, 
it will be obvious that any close-to-normal anchoring angle will lead to 
exactly the same situation,
as long as the director does not become almost tangential to the surface.
Another important point here is the size of the object. For extremely tiny
objects, the gradient energy of the winding may be too large so that
it may become energetically favorable for the  order parameter to assume
small (or vanishing) value at the surface. In that case no winding number
will be generated at the surface and one will not expect any brushes to
result. Such energetic balance is precisely what determines the
core size of a topological defect. Thus, one will expect that as long as
the size of the object is bigger than the typical size of the core of
a topological defect, one will expect a normal (or close to normal)
boundary condition for director anchoring  to lead to non-trivial
brushes. One can then use these brushes to detect presence of such a foreign
object in the liquid crystal sample. 

 The above discussion also suggests the following possibility. Consider a
foreign object of solid cylindrical shape immersed in an NLC sample. Assume
normal boundary conditions (for simplicity) for the director at the surface of
this object. The surface of solid cylinder is homeomorphic (topologically
equivalent) to the surface of a 2-sphere  (S$^2$). Thus one expects the 
same topology of four dark brushes originating from this object with
the cross-polarizer set-up. However, the geometry/shape of these brushes
need not be the same as for the spherical body case. In fact, for a
relatively long cylinder (compared to the diameter of the circular 
cross-section) one expects some sort of flat bands to originate from the 
surface along the length of the cylinder, while sort of conical brushes to 
originate from the circular disks at the ends of the cylinder, assuming
that diameter d of the circular cross-section is much smaller compared to the 
length L of the cylinder. This shape asymmetry should become more pronounced 
for cylinders with larger aspect ratio (L/d). 
If this shape asymmetry survives for long distances which can be seen by 
optical microscope, then even for object of few tens to few hundred nanometer 
sizes, the aspect ratio of the object may be determined using simple optical
microscopy. This can allow for easy identification of various microbes
or at least narrow down various possibilities. For certain microbes, this 
may even lead to non-trivial topology of brushes. For example, for
spiral shape bacterias, or viruses (e.g. Filoviridae viruses, such as the 
Ebola viruses, which form long, sometimes branched, filaments of varying 
shapes, with very large lengths of few thousand nanometers, with a diameter 
of about 100 nanometer) resulting brushes may show non-trivial patterns
depending on the twists along the length of the microbe. Again, such patterns
may allow easy identification of such microbes with optical microscopes.
 
 We explore this possibility in this work using numerical simulations.
Our results support the above mentioned possibility. We study brushes
emanating from cylindrical shapes of different aspect ratios and study the
resulting shapes of brushes. We find that the geometry of brushes retains 
some information about the aspect ratio of the cylindrical body to
significant distances beyond the surface of the microbe, even when different 
dimensions of the body are in few tens to few hundred nanometer
range.  We also study structure of brushes emanating from spherical bodies
of different radii exploring the possibility if the diameter of the sphere
could be identified from optical observation of brushes. This could then
help in identifying, or narrowing down possibilities, for a corona virus
(though, we mention that for roughly spherical microbes, complete symmetry 
of brushes itself is a useful information as it can help 
distinguishing from general asymmetric impurities/microbes). 
However, we find that in this case the information about the radius
of the spherical body is not reflected in the brushes at long distance,
with brushes looking the same for a large range of radii, (e.g. from 30 
nanometer to about 150 nanometer). We then studied the structure of brushes 
in the presence of electric field, with brushes folding down to two domain 
wall sheets emanating 
from the spherical body.  This set-up shows some promise of determining the 
size of spherical body looking at close-distance observations, though it is 
not clear if optical microscope will be able to 
clearly identify such patterns. One
possibility remains to be explored is whether the spikes on the surface
of corona viruses can induce some non-trivial patterns in the brushes at 
long distances, allowing for easy identification. This possibility is very
hard to explore numerically. It will be very useful if this possibility can
be explored directly by carrying out experiments with immersion of
corona virus in liquid crystalline samples. One important point we mention is
that we carry out these simulations using parameters for a
specific thermotropic nematic liquid crystal system MBBA (having critical
temperature $T_c^*$ for isotropic-nematic (I-N) transition  about 47.0 $^0$C).
For identification of microbes, e.g. corona virus, or bacteria, these
microbes need to be immersed in the liquid crystal sample. It may be much 
more convenient to  use lyrotropic liquid crystal \cite{lyrotropic} as  
immersing of microbes will be easier in an aqueous solution. We expect that
the qualitative aspects of our results will remain the same for 
lyrotropic liquid crystals also. With this note, 
we will continue to use thermotropic MBBA parameters for the simulation.

 We mention that there have been studies where effects of microbes on
liquid crystal ordering has been investigated (in lyrotropic liquid crystals 
as well as micro droplets of thermotropic liquid crystals) 
\cite{nlcmcrb1,nlcmcrb2,nlcmcrb3,nlcmcrb4}. (See, ref. \cite{nlcmcrb5} for 
brief reviews of investigations of liquid crystal based biosensors.)
However, there typically a finite 
concentration of microbes has been considered and its effects on liquid 
crystal ordering has been discussed. For example, in \cite{nlcmcrb1},
it is observed that different types of bacterias and viruses can
lead to transition from a bipolar to radial configuration of ordering
in droplets in aqueous emulsions of thermotropic liquid crystals.
Dependence of this bipolar to radial transition on
the droplet size, pH etc. has been investigated in \cite{nlcmcrb2}. 
Detection of E. coli strains using nematic liquid crystal 5CB with
crossed polarizer has been reported in \cite{nlcmcrb3}.
For lyrotropic liquid crystals, it was reported in \cite{nlcmcrb4}
that the presence of pathogens could be detected as optically bright
regions arising from distortion of the director around the immune 
complexes in a crossed polarizer setup. As we mentioned, in all these
studies, a finite concentration of microbes has been considered and its 
effects on liquid crystal ordering has been discussed. To our knowledge, 
use of geometry of brush structure emanating from a single microbe for 
possible identification of microbes has not been discussed previously.

  The paper is organized in the following manner. In Section 2, we write
down the order parameter and the Landau-de Gennes free energy  to be used for
numerical simulation. Section 3 discusses numerical simulation procedure.
As we are interested in static order parameter configuration, we use free 
energy minimization technique for determining the order parameter configuration
with given boundary conditions and the structure of resulting dark brushes.
Section 4 presents some specific test simulations to see that the algorithm
is producing results as expected. Here we show results of simulations of 
phase transitions which result in production of network of topological
defects. By suitably changing the NLC order parameter space to S$^1$, and
S$^2$, respectively we show that string network and monopole networks
are formed respectively for the two cases as expected. We then study the
case of RP$^2$, the NLC order parameter space and show that it results in
string network along with few monopole like structures, as expected. Section 
5 discusses
the case of microbes. Here we describe the boundary conditions for simulation 
of resulting brush structure.  We first consider cylindrical bodies and show 
that large distance brush structure reflects the aspect ration of the 
cylindrical shape of the body. We then present results for brush structures 
for spherical bodies of different radii, concluding that information of radii
is not reflected in large distance brush structure. We then study the brush 
structure for spherical bodies in the presence of electric field which
leads to brushes folding to form domain walls (sheets) emanating from
the core. We show that the close distance behavior of these sheets near the 
spherical core seems to reflect some information about the radius of the sphere.
We conclude in Section 6 with a discussion of certain important issues 
regarding our investigation. We discuss various implication of our results 
and speculate about possible brush structures for a body like corona virus 
with spike proteins.

\section{The order parameter and Free energy}

 For uniaxial nematic, we use the tensor order parameter 
\cite{priestley}

\be
Q_{ij}({\vec r}) = {1 \over 2} S({\vec r})[3n_i({\vec r}) n_j({\vec r}) 
- \delta_{ij}]
\ee

Where $S$ is the uniaxial order parameter and ${\hat n}$
is the nematic director. We use the following expression for
the Landau-de Gennes free energy density with terms
up to second order in $Q$ \cite{priestley}

\begin{align}
F = 
& \, {1 \over 2} a(T-T_c^*) Tr(Q^2) + {1 \over 3} B Tr(Q^3)
+ {1 \over 4} C (Tr(Q^2))^2 + \nn \\ 
& \, {1 \over 2} L (\partial_i Q_{jk}) (\partial_i Q_{jk})
\end{align}

We have not used here kinetic terms for general distortions.
In terms of $S({\vec r})$ and ${\hat n({\vec r})}$, $F$ can 
be written as

\begin{align}
F = 
& \, {3 \over 4} a (T-T_c^*) S^2  + {1 \over 4} B S^3
+ {9 \over 16} C S^4 + \nn \\ 
& \, {3 \over 4} L (\grad S)^2 +
{9 \over 4} L S^2 (\partial_i n_j) (\partial_i n_j)
\end{align}

We will use following values of parameters for MBBA \cite{priestley}.

\begin{align}
& \, a = 0.042 \times 10^6 {N \over K m^2} \nn \\
& \, B = -0.64 \times 10^6 {N \over m^2} \nn \\
& \, C = 0.35 \times 10^6 {N \over m^2} \nn \\
& \, L = 1.53 \times 10^{-12} N \nn \\
	& \, T_c^* = 47~ {^oC}
\end{align}

 For the numerical simulation purpose we change the length scale to
 nanometers. With that, all the parameters of the free energy become
 of order one with an overall factor of $10^{-12}$. All the plots
 will be shown in nanometer length scale.

 We will start the simulation with distribution of $S$ and ${\hat n}$
 at lattice points with appropriate boundary conditions. Specific
 boundary conditions will be used for the sample boundaries as well as
 for the surface of the supposed microbe under consideration.  The above 
 free energy will be minimized and for the resulting order parameter
 configuration (for $S$ and ${\hat n}$) resulting dark brush structure 
 will be determined assuming a cross-polarizer setup.  As we mentioned 
 in the Introduction, for spherical microbe structures, the large length 
 scale structure of dark brushes is not able to give information
 about the radius of the sphere (for radius in the range of few tens
 nanometers). To probe the size for the spherical case, we have also
 considered the effect of applied electric field which aligns the 
 director field everywhere. With constraints of topological invariance of 
 winding of the director, this leads to expansion of two brushes (for the
 electric field along or perpendicular to the polarization axis) while 
 two other brushes fold to form domain wall sheets emanating from the 
 spherical core. The thickness of this sheet is determined by the strength of 
 the electric field, with stronger electric field leading to thinner sheet.
 Thus, when the size of the spherical body becomes comparable
 of larger than the sheet thickness, the sheet will undergo bulge like 
 deformation near the spherical core, indicating presence of the spherical 
 body. This bulging in some sense slightly magnifies the spherical core,
 increasing chances of its detectability. If this can be detected
 then it can be very useful in restricting possibilities for the 
 spherical body, from general impurity to a body of specific size, say 
 a corona virus. (As many microbes are of sizes just below threshold
for detectability using an optical microscope system, even a small 
magnification of the effective core region can be useful.)

 We consider electric field applied along the $y$ axis, so ${\vec E}
 = E_y {\hat y}$.  We will consider cross-polarizer setup in the x-y plane 
 with light propagating along the $z$ direction for studying the brush pattern.
 With the electric field contribution, we use the following free 
energy \cite{efield}

\be
F_{E} = F - {1 \over 2} \epsilon_0 \Delta \epsilon 
({\vec E}.{\vec n})^2 = F - S E_0 n_y^2
\ee

In writing the right-most term, we have taken the dielectric constant to be  
proportional to the  order parameter $S$, \cite{rfrctindx}, with $E_0$ 
representing rest of the factors (including $E_y^2$). Here,
$F$ is the free energy (Eqn.(3)) in the absence of any
external field (we have ignored here a contribution from the electric 
field which does not depend on the orientation of ${\vec n}$ as our
main motivation is to find structure of dark brushes resulting from
orientational changes of ${\vec n}$).  $\Delta \epsilon$ is the 
difference between dielectric constants of extra-ordinary and ordinary rays. 
We will consider a sample value of the factor $E_0$ = 0.01. For this value, 
we will consider cores of different radii, and see if for some values 
observations of brushes can indicate the size of the spherical core.

\section{Numerical simulation}

 We consider a suitable lattice size, depending on the specific case under
consideration, and prescribe values of the NLC order parameter field,
the scalar order parameter $S({\vec r})$ and the director ${\hat n}({\vec r})$
at each point of the lattice. Suitable boundary conditions, again
depending on the specific case, are prescribed for $S$ and ${\hat n}$.
As we are only interested in static configurations, we follow the
procedure of free energy minimization to determine the correct
nematic order parameter field configuration with given boundary conditions.

 In the following we list main steps of the algorithm for the energy 
 minimization in our numerical simulation procedure.

\begin{itemize}

\item  
A suitable initial  order parameter field configuration is prescribed
on the three-dimensional lattice (lattice size and lattice spacing
are appropriately chosen for the specific case being considered,
as mentioned in the respective sections below).
This amounts to specifying a suitable initial value for $S$
at each site, and the polar angle $\theta$ and the azimuthal angle 
$\phi$ for the director ${\hat n}$. 

\item 
Each field variable, namely $S$, as well as the orientation of the director
$\theta$ and $\phi$, are fluctuated by small amounts (say, for $\delta S$, 
typically 5 to 10 \% of the starting value of $S$, somewhat smaller
initial fluctuations are found appropriate for $\theta$ and $\phi$) and the 
free energy (Eqn.(3)) is calculated for the three values, for the starting 
value $S$, for $S + \delta S$, and for $S - \delta S$ (similarly for 
fluctuated $\theta$ and $\phi$ values). 

\item
We use over relaxation technique for determining 
the final fluctuated configuration at each step. (We have used this
technique for different field simulations and have found it to be very
efficient  \cite{overshoot}.) This consists in first 
determining the most favorable fluctuation in the field (say, $S$) at a 
given site by fluctuating $S$ there and considering the change in the 
free energy density. The most suitable fluctuation corresponds to the 
minimum of the parabola which passes through these three values of energy
densities (corresponding to fluctuated values of $S$). Then the actual 
change in $S$ is taken to be larger (by a certain factor) than this most 
suitable fluctuation.  We have found that changing this {\it overshoot} 
factor in the range 0.02 - 0.2 worked best for our case.

\item
 For the nematic order parameter case, at each step of the above 
free energy minimization, we ensure that
in comparing a fluctuated value of the free energy, say $F_1$ to
the starting value $F$, we calculate two values of $F_1$, one with
$+{\hat n}$, say, $F_1^+$ and the other with $-{\hat n}$, say, $F_1^-$.
This is because of the identification of ${\hat n}$ and $-{\hat n}$
for the nematic  director. The lower value among the two values
$F_1^+$ and $F_1^-$ is then accepted as the correct choice.

\item
 This minimization technique works fine with free boundary conditions, or
with fixed boundary conditions when $S$ and ${\hat n}$ are completely
fixed at the boundaries. However, the director is known to anchor
at specific angles at certain specific boundaries, which differs from the
normal anchoring \cite{angle}. In 
that case the director lives on a cone with the
half angle of the cone being given by the anchoring angle of ${\hat n}$
at that specific boundary. In the absence of knowledge about the anchoring of
${\hat n}$ at the surface of the specific microbe under consideration, we
have also considered such boundary conditions. In such a case, the polar
angle of ${\hat n}$ (determined w.r.t the local normal to the surface of
anchoring) is fixed, but the respective azimuthal angle is allowed full
variation during free energy minimization.

\item
 Minimization procedure is continued until change in the free energy becomes
insignificant. Often, when such a state is achieved, changing the 
magnitude of fluctuation ($\delta S$, $\delta \theta$, $\delta \phi$),
some times decreasing it, some times even increasing it, again starts
giving significant changes in the free energy showing that earlier
minimization led to a local minimum of the free energy, and larger fluctuations
were needed to hop over some barrier to explore a new, lower minimum.

\end{itemize}

From the last step of the algorithm above it is clear that one is never 
sure if the final configuration indeed represents the {\it correct} order 
parameter configuration with the lowest possible free energy. Precisely for
this reason, we have carried out different tests of our minimization code,
as we discuss in the next section, which gives results as expected from
general known simulations about the defects. This gives us greater
confidence in our final results. Of course, the final check on these
results has to come from experimental observations.

 Our main proposal is to be able to identify the microbe shape using
optical microscopy with cross-polarizer set up. Thus, we have to determine
the geometrical shapes of the dark brushes expected in these cases.
We consider plane polarized light of specific wavelength $\lambda$ 
propagating from the bottom of the cell to the top of the cell along a given 
axis (z axis, or x axis, or y axis, depending on the specific shape of the 
central object being probed). The polarization of this light is 
decomposed in two components at each lattice site depending on the
local direction of the director ${\hat n}$. Each component then
propagates in the z direction with respective speed
depending on the corresponding refractive index. This leads to rotation
of the polarization vector for a general orientation of ${\hat n}$ as the
light propagates from the bottom of the sample to the top.

 For getting clear picture of brushes, we calculate the polarization 
rotation only for small thickness of the sample in middle, consisting
about 20 lattice points. Also, we use sample values for the refractive
indices for the NLC. The aim here is not to calculate exact appearance of
the dark brushes for the sample under consideration, but get the correct
geometry of the brushes. Thus exact contrast of brushes to the background
is not relevant.

For light traveling along the $z$ direction (when brushes are 
viewed in the x-y plane), we take $\Delta z$ to be the length travelled by 
the light.  For this case, the starting polarization is taken to be 
 along the x-axis and a crossed polarizer is taken on top, along the y axis.
(Similar arrangements are done for light traveling along x axis, or along 
y axis).  The resulting optical intensity is taken to be proportional 
to \cite{intensity},

\begin{equation}
I = I_0 sin^2(2\phi)sin^2(\pi \Delta z \Delta r/\lambda)
\end{equation}

Here, $\Delta r = (r_e - r_o)sin^2\theta$, $r_e$ and $r_o$ being refractive 
indices for the extra-ordinary and ordinary rays respectively and $\theta$
is the polar angle (for light traveling along $z$ axis case). 
$\phi$ is the average director orientation from x axis for
the particular light ray under consideration \cite{intensity}.
$\lambda$ is the wavelength of the light. 
Taking the dielectric constant to be  
 proportional to the  order parameter $S$, we use following dependence
 of the refractive indices on $S$ \cite{rfrctindx}.

\be
{\epsilon_n (S) \over \epsilon_0} = r_n(S)^2 = [r_i^2 - 
(r_i^2 - r_n^2)({S \over S_0})]
\ee

Here, $n$ subscript denotes either ordinary ray or the extra-ordinary ray,
so $r_n$ is $r_o$ or $r_e$, and $r_i$ is the refractive index for the
isotropic phase. $\epsilon_0$ is the permittivity of the vacuum.
$S_0$ is the equilibrium order parameter for the given temperature and 
$r_n(S)$ is the refractive index with value of order parameter being $S$. 
We use sample values of these parameters,  $\lambda = 600 $nm, 
$r_e = 1.75, r_i = 1.62$ and $r_o = 1.55$ \cite{rfrctindx}.
We have also varied these parameters, as well as the thickness 
$\Delta z$ for calculating brush structure. As expected, even though
contrast with background changes, the geometry of brushes remains
unaffected.

\section{Tests of the simulation technique}

 As we mentioned above, getting correct order parameter configuration by
energy minimization is complicated. Even when free energy appears minimized
with subsequent fluctuations making insignificant changes in the free energy,
changing the magnitude of fluctuation (decreasing it, or some times even 
increasing it), again starts giving significant changes in the free energy 
showing that earlier minimization had led to a local minimum of the free 
energy, 
and larger fluctuations were needed to hop over some barrier to explore a new, 
lower minimum. Thus, one needs to develop confidence that the minimization
technique is giving desired result, at least qualitatively. For this
purpose, we  now discuss different tests of our minimization code which we
have carried out. These produce results as expected from general known 
simulations about the defects giving us greater confidence in our final 
results.

 The tests we carry out here are for phase transition dynamics and we
determine qualitative structure of the resulting defect network. This is a 
comprehensive test as in this case the full order parameter, the scalar 
part $S$, as well as the director ${\hat n}$, undergo large variations 
during energy minimization, and the coarsening of resulting defect networks
is observed. If this produces desired results for the defect network and its
evolution, we have greater confidence in the main simulations where main focus
is only on the brush structure emanating from a central core object.
We will study cases corresponding to different
order parameter spaces by suitably choosing the degrees of freedom of
director ${\hat n}$. We consider the cases of string production with
$S^1$ order parameter space, monopole production with $S^2$ order
parameter space, and strings and monopole network production with $RP^2$
order parameter space as appropriate for NLC case.

 We emphasize that we are not attempting to probe  quantitative aspects of
defect network which results in a phase transition. The network we
get is highly non-equilibrium with underlying mechanism of formation being
the so called {\it Kibble Mechanism} \cite{kbl}. 
In this case a dense network of
defects is formed at the phase transition which coarsens rapidly eventually
leading to homogeneous phase (apart from isolated defects pinned to boundary).
To get such an evolving network one needs to carry out simulation using
full time-dependent Landau-de Gennes equations. Our main purpose is in the 
static order parameter configuration in presence of specific boundary 
conditions, specified at the sample boundary, as well as on the surface of
the microbe. For this we are using energy minimization technique. Thus,
for the phase transition case, we will get qualitatively correct network,
after all highly dissipative dynamics is like free energy minimization.
However, different steps of simulation during the minimization of free energy 
will not have any direct correspondence to actual time steps. We will see
that the defect networks we get have correct qualitative behavior. We get
a network of defects as expected for the relevant case, which coarsens as
free energy is further minimized.

 We start the simulation by starting with $S = 0$ with tiny fluctuations
$\delta S$ and random variations of ${\hat n}$ within the 
allowed degrees of freedom for the respective case under consideration. 
This will represent the situation of the isotropic phase, with fluctuations
representing thermal fluctuations around the equilibrium order parameter for
the isotropic phase. We set temperature to a value less than the
critical temperature so that with energy minimization, the order parameter
settles to a non-zero value as appropriate for the nematic phase.
As the isotropic-nematic phase transition is weakly first-order, we
set the temperature to a suitably low value so that there is
no local minimum at $S = 0$ (otherwise energy minimization may not
take the order parameter outside this local minimum, remaining trapped within
the free energy barrier). This is easily achieved by setting $T = T_c^*$.
With this the $S^2$ term vanishes and there is no metastable state at 
$S = 0$.  The situation here therefore corresponds to a quench.
Free energy minimization is carried out and
field configuration is plotted at different stages during the minimization.
 
  For all the plots shown in the figures in this section, the defect core
is shown by making a density plot with suitable cutoff in the density.
We have used energy density plot as well as plot of $S_0 - S$ ($S_0$
being the equilibrium order parameter corresponding to minimum
value of the free energy for a uniform order parameter). Both plots show
similar defect network. Also, to show defect cores clearly, we have
generally plotted smaller portion of the lattice. Periodic boundary
conditions have been used for the lattice boundary.

\subsection{String network with $S^1$ order parameter space}
 
 To get $S^1$ order parameter space, we restrict the polar angle of 
${\hat n}$ to $\pi/2$. Further, in the energy minimization, we drop
the step where we impose the identification of ${\hat n}$ with $-{\hat n}$.
This reduces the original $RP^2$ order parameter space to effectively
$S^1$ case (same as for the $X-Y$ model). The only topological defects in
this case are string defects (in 3-D).

 Fig.1 shows the resulting string network. The thickness of the tube
representing strings depends on the choice of cutoff we impose for
plotting energy density (or the value of $S_0 - S$). We have chosen
a value suitable to clearly see the string network as well as the core
of the string (shown by the color contrast in the cross-section of
tubes which is visible at the boundary of the plot). We have confirmed 
that these are genuine string defects  with plots of ${\hat n}$ around
these strings which clearly shows non-trivial winding of ${\hat n}$.
As mentioned above, we show a smaller part of the whole lattice used 
for simulation.  we have used lattice of size $50 \times 50 \times 50$ 
with lattice spacing of 10 nm.

\begin{figure}
\centering
\includegraphics[width=0.6\linewidth]{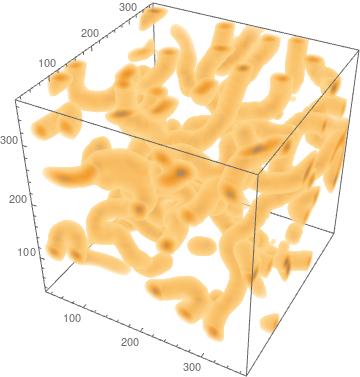}
\caption{}{String network resulting in the quench. String profile
can be seen at the cross-section of the core at the boundary. Initial
string network is extremely dense.  Length is given in nm.  We show the
situation after suitable number of minimization steps so that the 
string network has sufficient coarsened and individual strings can be
seen. }
\label{fig:fig1}
\end{figure}

 This string network has correct qualitative structure as obtained in
various simulations of $S^1$ string defect formation (from relativistic
cases to non-relativistic ones), basically representing a random
walk network with the formation of small string loops as well as many 
long strings. With further minimization of free energy, the network is
seen to coarsen further leading finally to only 1-2 strings with ends
pinned at the boundaries. Note that with periodic boundary conditions,
one is working with the topology of a 3-torus $T^3$ for the physical 
space. A single string winding around any of the circles of $T^3$ is 
topologically stable and cannot be removed.
  
\subsection{Monopoles with $S^2$ order parameter space}

 Monopoles also arise with $RP^2$ order parameter space of NLC. However,
In that case one also gets strings. To clearly see monopole distribution, it 
is more convenient to consider the  more standard case of $S^2$ order parameter
space where only monopole defects are formed.To get $S^2$ order parameter 
we only drop the step of identification of ${\hat n}$ with $-{\hat n}$
in energy minimization.  This reduces the original $RP^2$ order parameter 
space to effectively $S^2$ case.

 Fig.2a,b show the resulting monopole distribution network. Each blob shows
a monopole with the size of the blob depending on the choice of cutoff we 
impose for plotting energy density (or the value of $S_0 - S$). Fig.2a shows
a relatively large part of the lattice while Fig.2b shows a smaller
portion showing clearly isolated monpoples. It is
important to mention that several such localized blobs are seen after
the quench which individually disappear with energy minimization. Such
blobs represent only local fluctuations without any non-trivial topological 
monopole winding number associated with them. The blobs which survive
are energetically stable and represent genuine monopole configurations.
We have confirmed this with the plots of ${\hat n}$ around such blobs
clearly showing non-trivial winding of ${\hat n}$.

\begin{figure}
\centering
\includegraphics[width=0.9\linewidth]{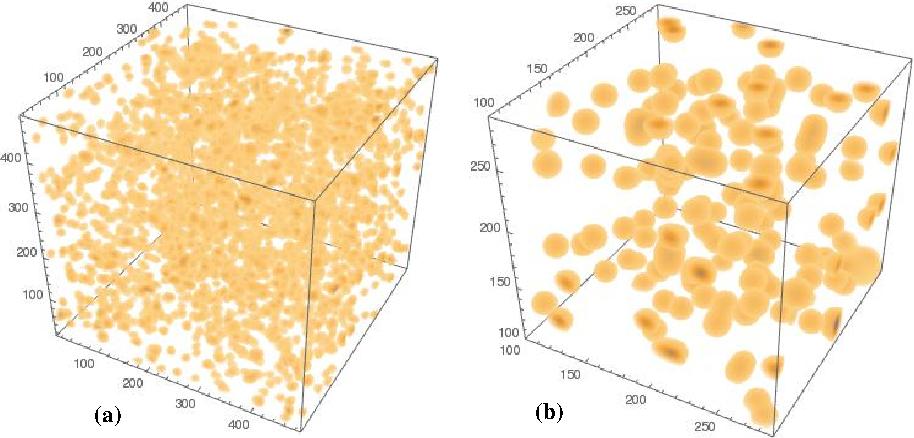}
\caption{}{Monopole distribution resulting in the quench with $S^2$
order parameter space. (a) shows a relatively large part of the lattice 
while (b) shows a smaller portion showing clearly isolated monpoples.
Length is given in nm. }
\label{fig:fig2}
\end{figure}

\subsection{Defect network with NLC order parameter space $RP^2$}

 We now consider the case of full $RP^2$ order parameter space of
NLC. ${\hat n}$ is allowed full variations of the polar angle $\theta$
and the azimuthal angle $\phi$. Further, identification of ${\hat n}$ 
with $-{\hat n}$ is imposed  in energy minimization as explained
in the previous section. We expect string network along with few
monopoles. This is what is seen in Fig.3a,b. Fig.3a shows the network
for a relatively large portion of the lattice while Fig.3b shows a zoomed
portion to clearly show an isolated blob corresponding to the monopole.
It is known that number of monopoles is highly suppressed
for $RP^2$ case when monopoles are formed directly via the Kibble
mechanism \cite{monopole}. However, monopoles can also form by the 
collapse of a string loop in this case. The few monopoles we
see could well be result of collapse of  small string loops.

\begin{figure}
\centering
\includegraphics[width=0.9\linewidth]{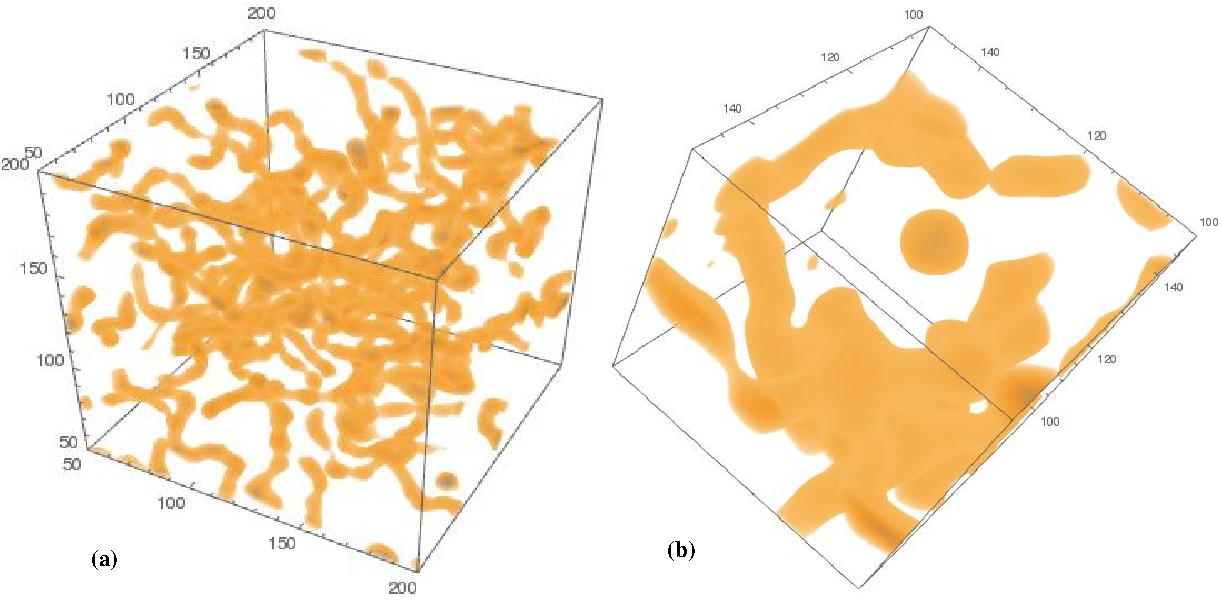}
\caption{}{(a) Defect distribution for the $RP^2$ order parameter space 
of NLC.  String network and few isolated monopoles can be seen. (b) shows
zoomed portion clearly showing an isolated monopole blob.
Length is given in nm. }
\label{fig:fig3}
\end{figure}

 These results for the defect networks for different order parameter
cases are in confirmation of the results in literature from dynamical
simulations of phase transition. For example, pictures in \cite{network}
show similar defect network where molecular dynamics simulation was used 
to model I-N quench.

\section{Probing microbe shapes with the geometry of dark brushes}

 With the results in previous section in confirmation with the usual
expectations for defect networks, we now move on to our main goal, to
determine geometry of dark brushes for different shapes of microbes.
We discuss two classes of microbes here. One with cylindrical shape,
and the other with spherical shape which is the case relevant for
corona virus. First we consider the cylindrical shape case.
We have used lattice size $150 \times 150 \times 150$
and used spherical boundary with director
anchored normal to the boundary.  The microbe is situated at the centre of
the lattice. Lattice spacing is taken to be 10 nm.
Due to computer limitation, we are able to use relatively small lattice
size here of order of few microns. This is sufficient for our purpose, as
we imagine a spherical micro droplet of the sample being used for the
microbe detection purpose which should be detectable using optical microscope.

\subsection{microbes with cylindrical shape}

 We will consider shape of the microbe in the form of a solid cylinder, with
flat disk shape at the end of the cylinder. Cylinder is kept with its length 
along the z axis, centered at the origin.  We use normal boundary conditions 
for the surface of the cylindrical microbe. It is then useful to have a 
healing length for the order parameter so that ${\hat n}$ is kept fixed 
within the healing length from the 
surface.  After that its orientation is allowed full fluctuations to determine
the correct ${\hat n}$ distribution. For all the cases here we have taken
healing length to be one lattice spacing.  With this orientation
of the cylinder, if the cross-polarizer setup is in the x-y directions 
respectively, four dark brushes are expected emanating from the surface of
the central body (which will have a circular cross-section in the x-y plane), 
two along $\pm$ x axis and two along $\pm$y axis. Similarly, if the 
cross-polarizer setup is in the y-z directions respectively, four dark 
brushes will emanate from the surface of the central body (which now
will have a rectangular cross-section in the y-z plane), 
two along $\pm$ y axis and two along $\pm$z axis. 

 We carried out simulations in two different manners. First we started with
a reasonable guess of director ${\hat n}$ distribution, with the order 
parameter $S$ having the equilibrium value, and allowed only the director  
orientation to relax by energy minimization. We found reasonable results,
however, the memory effect of initial choice could not be completely 
eliminated. To be completely sure of our results, we carried out the quench 
transition as in Section 4,  by starting with scalar order parameter $S$ to 
be zero, with tiny random fluctuations in magnitude of $S$, and in the 
orientation of the director. However, at the boundaries
(outer boundary as well as on the microbe surface) we take equilibrium
value $S = S_0$ with normal anchoring of the director. During energy
minimization, a defect network is formed which rapidly coarsens, and finally
leads to the director configuration appropriate for the given boundary 
conditions.  We show in Fig.4, a sequence of pictures from the simulation 
showing different stages during energy minimization.  Here, the central 
object is taken to be of cylindrical shape with diameter of 60 nm and length 
300 nm, with its length along the z axis.  (a)- (d) show time sequence of 
brushes viewed along the z axis during the coarsening, starting from
the configuration at beginning stages in (a), with (d) showing the final
brushes with fully relaxed order parameter configuration. (e)-(h) show time
sequence of brushes viewed along the x axis from an intermediate stage
in (e) to the final stage in (h). Shape of the central cylindrical object is 
shown by the black solid circle in (d) and black solid rectangle in (h). 
Symmetry of brushes in Fig.4d and the asymmetry of brushes in Fig.4h is clearly 
seen, directly originating from the corresponding shapes of the central object.

\begin{figure}
\centering
\includegraphics[width=1.0\linewidth]{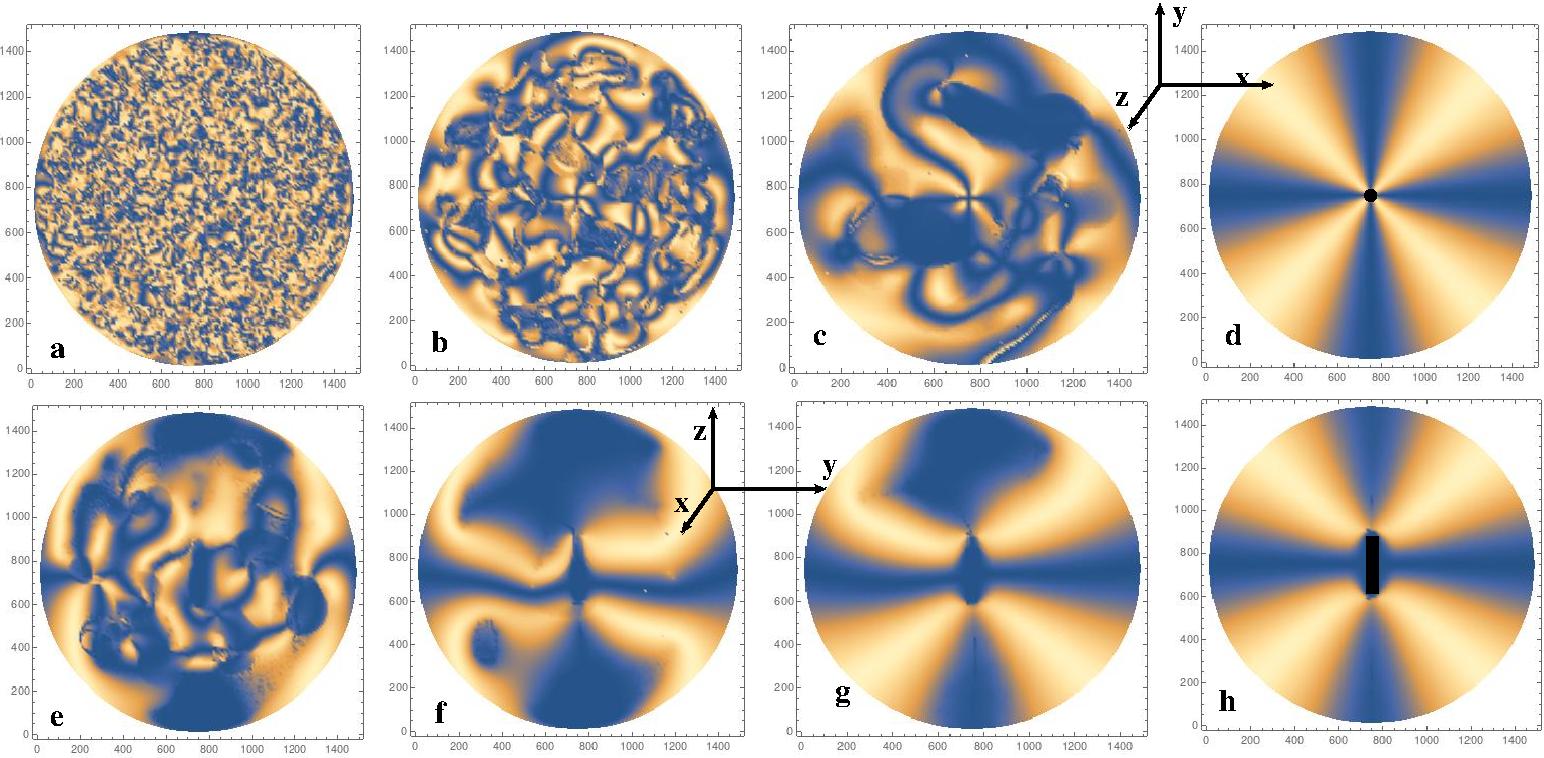}
\caption{ A sequence of pictures from a simulation showing different
stages during energy minimization for a quench with fixed boundary conditions
at the outer spherical boundary as well as the boundary of
central object taken to be of cylindrical shape with diameter of 
60 nm and length 300 nm, with its length along the z axis.  (a)- (d) show time
sequence of brushes viewed along the z axis during the coarsening, starting 
from the configuration at beginning stages in (a), and (d) showing the final
brushes with fully relaxed order parameter configuration. (e)-(h) show time
sequence of brushes viewed along the x axis from an intermediate stage
in (e) to the final stage in (h). Shape of the central cylindrical object is 
shown by the black solid circle in (d) and black solid rectangle in (h). 
Symmetry of brushes in (d) and the asymmetry of brushes in (h) is clearly 
seen, originating from the corresponding shapes of the central object.}
\label{fig:fig4}
\end{figure}

The final brush configuration obtained by this method was found to be very
robust, using different lattices, different lattice spacing  etc. We thus
used this quench technique to determine the brush geometries for all the
cases for the cylindrical shape case.  Fig.5 shows three different cases.
As we are only interested in the geometry of brushes, we have used a light
intensity cut-off (with the  cross-polarizer set-up) so that only brushes 
are seen in the figures. These brushes correspond to the configuration 
similar to the one shown in Fig.4h. Thus, as the cylinder length is along the
z axis, and these figures correspond to light propagating along the x axis, 
the brushes shown here are in the y-z plane.  Figs.5a,b,c show the cases 
of cylindrical cores with (diameter,length) being (100 nm, 300 nm),
(60 nm, 300 nm), and (60 nm, 420 nm) respectively. Brush edges are marked by 
dashed lines and corresponding angles $\alpha, \beta$ are measured (by 
transporting one of the lines towards the other line for each brush).
$\beta_1,\beta_2$ angles correspond to the brushes emanating along the y axis
while $\alpha_1,\alpha_2$ angles correspond to brushes emanating along the
z axis.  $\alpha, \beta$ angles used (in Table 1 below) are the averages of the
angles $\alpha_1,\alpha_2$ and $\beta_1,\beta_2$ respectively in Figs.5a,b,c.

\begin{figure}
\centering
\includegraphics[width=1.0\linewidth]{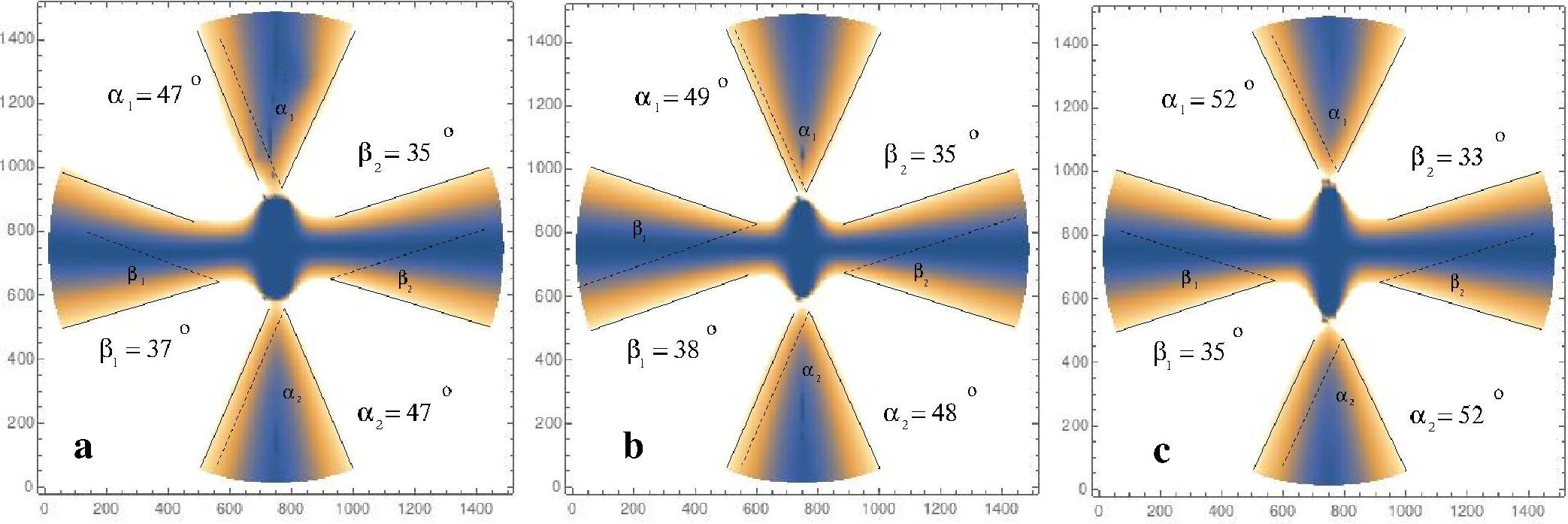}
\caption{Brushes (for the final configuration) are shown here by using a 
cutoff for light intensity under cross-polarizer.  (a),(b),(c) show the cases 
of cylindrical core with (diameter,length) being (100 nm, 300 nm),
(60 nm, 300 nm), and (60 nm, 420 nm) respectively. Brush edges are marked by 
dashed lines and corresponding angles $\alpha, \beta$ are measured (by 
transporting one of the lines towards the other line for each brush).
$\alpha, \beta$ angles used (in Table 1 below) are the averages of the
angles $\alpha_1,\alpha_2$ and $\beta_1,\beta_2$ respectively in figures
(a),(b),(c) above.}
\label{fig:fig5}
\end{figure}

Table 1 shows the results of measurements of brush geometries for different
cases. The shape of the brushes emanating along the y and z directions are
given by (average) angles  $\beta$ and $\alpha$. Results clearly show a 
monotonic increase of the ratio of the two brush angles $\alpha/\beta$ with 
increasing aspect ratio of the object. We do not attempt to get a quantitative
dependence of the angle ratio on the aspect ratio due to crudeness of our
results at this stage. We mention here that the detection
of brush edges here is done rather crudely, by visually drawing a straight
line along the brush edges which are not very sharp and often are not
of straight line shape.  Clearly a
more sophisticated technique of graphical analysis can give more reliable
results. Our intention here is to show a clear pattern of brush geometry
containing signature of shape asymmetry of the central core object.

\begin{table}
\centering
        \caption{Variation of brush angles with aspect ratio of the central
rectangular shape. Results clearly show a monotonic increase of the ratio of
the two brush angles $\alpha/\beta$ with increasing aspect ratio of
the object. Here, $\alpha$ and $\beta$ represent averages of the two angles
$\alpha_1,\alpha_2$ and $\beta_1,\beta_2$ respectively in Fig.5.}
\label{tab:table}
\begin{tabular}{lccccc} 
\hline
        Fig5. & cylinder size  & aspect ratio & angles & $\alpha/\beta$ \\
        lable & (diameter d, length L)  &  (L/d) & $\alpha, \beta$  &  \\
\hline
a & (100 nm, 300 nm) & 3.0 & 47.0$^o$,36.0$^o$ & 1.306 \\
\hline
b & (60 nm, 300 nm) & 5.0 & 48.5$^o$, 36.5$^o$ & 1.329\\
\hline
c & (60 nm, 420 nm) & 7.0  & 52.0$^0$, 34.0$^0$ & 1.529\\
\hline
\hline
\end{tabular}
\end{table}

\subsection{microbes with spherical shape}

 We now consider microbes of spherical shape. This case is of particular
importance as it relates to corona virus. Of course, the corona virus does
not have simple spherical surface, having spike proteins on the surface.
Thus the actual boundary conditions of NLC director may be much more complex
for corona virus. This has to be determined by experiments. We will ignore
such complications here and assume simple spherical surface with normal
boundary conditions for the director ${\hat n}$. This may not be unreasonable
as the spike proteins are relatively small, and NLC molecular ordering
in the nematic phase may be largely unaffected  by their presence.

 As our primary focus for the spherical case is regarding viruses, in
particular the corona virus, we keep the sizes relatively small
compared to the cylindrical case. For different radii we use different
lattice sizes as appropriate.  Fig.6 shows brush structure for widely different 
radii of the spherical microbe case (for light propagating along the z axis,
so the resulting brushes are in the x-y plane). 
(a) and (b) show cases with radii 
30 nm, and 150 nm respectively.  (Brushes appear somewhat 
different here compared to earlier pictures due to different color scheme in 
plotting.) It is clear that no difference can be seen at large scales in 
brushes which could indicate the size of the core. 
This is expected as with normal boundary conditions
on a spherical surface, brushes will appear to be originating from the
center, hence will all have the same shape at large scales.
With complete symmetry of brushes in x and y directions, one can
conclude that there is an almost spherical body present with normal boundary
conditions for the director, and that by itself is possibly some useful
information. (A general foreign body, or other microbes will in general not
have spherical shapes, and with our results shown in Figs.4,5, should
result in some asymmetry in brushes.) However, one would like to get some 
information about the order of magnitude of the size of the object.

\begin{figure}
\centering
\includegraphics[width=0.9\linewidth]{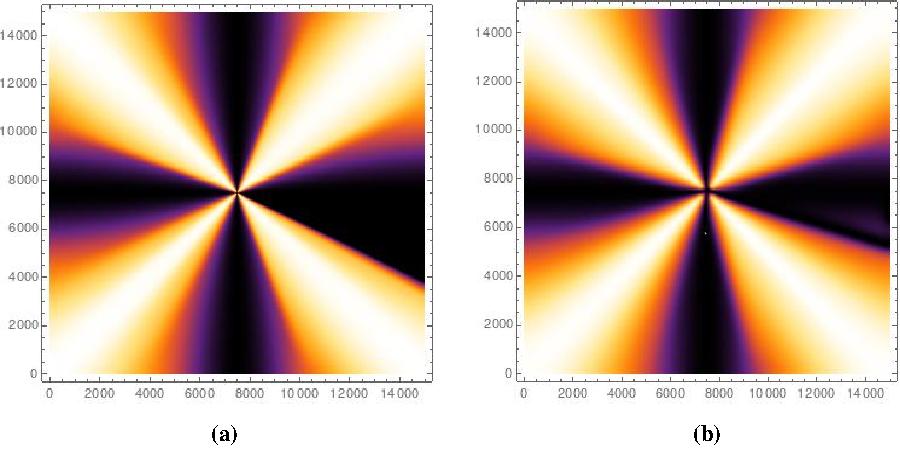}
\caption{}{Spherical microbe case with (a), (b)  showing
brush structures for spheres with radii 30 nm, and 150 nm,
respectively. No difference can be seen at large scales in brushes which 
could indicate the size of the core.}
\label{fig:fig6}
\end{figure}

 It is clear that the spherical symmetry of the microbe, with normal
anchoring of NLC director on its surface continues to dominate the large
scale brush structure, with the same brush structure for all radii
considered.  This also suggests that to probe the radius of the object with
normal boundary conditions, one should break this spherical symmetry.
This can be achieved by applying electric field, or magnetic field. As
an example, we consider application of electric field along y direction. 
The relevant fee energy density is given by Eqn.(5). We use $E_0$ (which
represents factors of dielectric constants, and $E_y^2$) as a variable 
without attempting to relate to actual value of the electric field 
applied. The lattice size is taken here as 500 $\times 500 \times 50$
with lattice spacing of 4 nm.

With electric field along y axis, the brushes along y axis expand while
those along the x axis fold to form domain wall sheets along the x axis
emanating from the spherical core. The sheets represent the regions where 
the director undergoes the winding as the folding of bands has to 
respect the topological invariance of winding arising from the boundary 
conditions at the surface of the sphere. For larger radius, the folding of
bands will extend to larger distances from the center. The thickness of 
the sheet is determined by the strength of the electric field $E_y$, 
with larger $E_y$ (hence larger value of factor $E_0$)leading to thinner 
sheet. When sheet thickness is
larger than the diameter of the spherical body, the presence of the body
will not be noticeable. However, when the size of the 
spherical body becomes comparable of larger than the sheet thickness, the 
sheet will undergo bulge like deformation near the spherical core as the 
director attempts to fold down, starting with normal anchoring at the
spherical surface, thus indicating presence of the spherical 
body and may become visible under optical microscope. Basically the electric 
field effect on folding the dark bands in form of domain wall helps in 
slightly magnifying the core size.  Thus, for a given applied electric field 
$E_y$ (with a value corresponding to choice of $E_0 = 0.01$ in Eqn.(5), 
including other factors), we 
consider spheres of increasing radii. For a suitable radius, the folding of 
bands should show up with some bulging up near the core which may become
visible under optical microscope even when the sphere size if few tens of
nm. Fig.7(a), (b), (c) show the cases of spheres of radii 28 nm, 52 nm, and 
152 nm respectively (specific choices being results of choice of lattice 
spacings).  We see that there is no change in the folded bands in (a) and
(b), but in (c) some bulging up can be seen near the core of the object.
This indicates some possibility that in suspected presence of spherical microbe
of some specific size (say, corona virus having radius of about 60 nm),
as one gradually increases electric field in a cross polarizer set up, some
bulges may start showing up at the locations of the microbes.  Proper 
calibration can relate that particular value of the applied electric field 
(when bulges start appearing firs) to the size of the spherical microbe.

\begin{figure}
\centering
\includegraphics[width=0.9\linewidth]{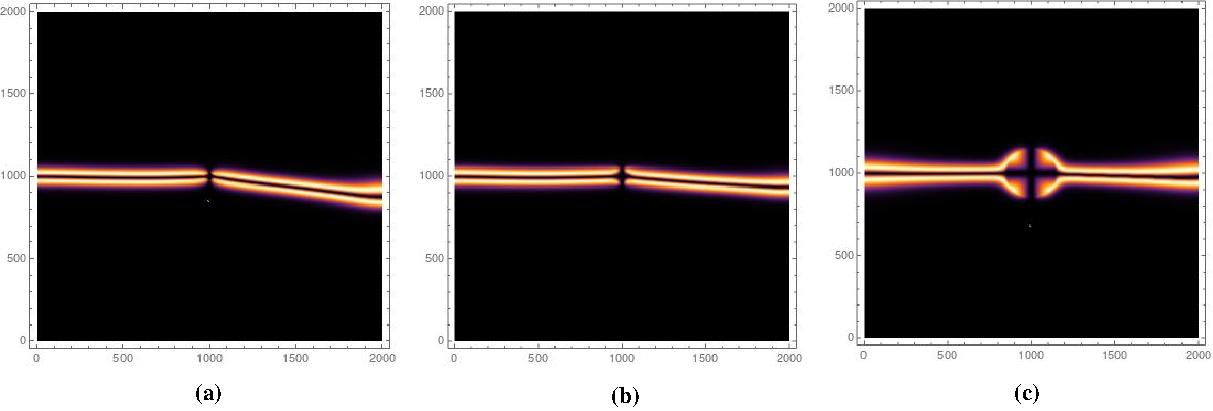}
\caption{}{Spherical microbe case in the presence of electric field of
fixed magnitude along y axis. (a), (b), (c) show folded brush
structures (due to electric field) for spheres with radii 28 nm, 52 nm, and 
152 nm respectively. No change in the folded bands is seen in (a) and
(b), but in (c) some bulging up can be seen near the core of the object.}
\label{fig:fig7}
\end{figure}

 We mention that all the results we have presented in this paper are with 
 normal boundary conditions. However, the results do not crucially depend
 on strict normal boundary conditions. We  have also allowed conical boundary 
 conditions for several cases, with half cone angle being about 30 degrees. 
 The large scale brush structure remains almost unaffected, simply being 
 determined by the geometry of the surface. This is expected as with
 conical boundary conditions, random variation in the azimuthal angle
 (w.r.t the cone axis) averages out in a given region of the surface,
 with average orientation being along the cone axis, i.e. normal to the
 surface. This is what seems to determine the overall geometrical structure
 of the brushes even with conical boundary conditions.

\section{Conclusions}

Our study here is aimed at providing possibilities. The main aim is to be
able to have a quick diagnosis using simple optical microscopy with 
cross-polarizers and narrow down possibilities of microbes which could be
present in the sample. As we mentioned, for microbes, it is more
appropriate to use lyrotropic liquid crystals. We use here specific case
of thermotropic  NLC. Qualitative aspects of our  results relating geometry
of brushes to the aspect ratio of elongated microbes, or the appearance of
bulges in the presence of specific value of electric field, should be
valid for lyrotropic case as well. We have only considered simple geometrical 
shapes for the microbes here which lead to standard 4-brush geometry
under cross-polarizer. Microbes occur in variety of shapes, including
complicated shapes, such as spiral shape bacterias or Filovirus such as 
Ebola viruse. In such cases, even the topology of brushes may be
non-trivial depending on the spiral nature/folding of the virus along
its length. In fact, one may expect linked/knotted brush structure
(with 2-D projection) which may uniquely identify the underlying
microbe structure. The important point here is that while other studies
probing microbes with liquid crystals involve clusters of microbes, here
one is attempting to observe single, individual microbe. Thus, even with
extremely low concentration of microbe, one may be able to detect it.
In fact, even if the microbe is not in the field of view of the microscope,
being slightly away, the brushes emanating from it may very well
be seen in the microscope. With the converging nature of the brushes
towards the microbe, one may be able to bring back the relevant region
under microscope for further detailed analysis.

  Our technique of determining the correct director configuration uses
energy minimization. One can never be sure with this method that correct
global energy minimum has been achieved. We have carried out different
tests of our technique by determining defect network structures expected
in a quench, and get qualitatively correct results. This gives us
the confidence that the qualitative behavior of brush structures we 
determine  should correctly represent the shapes of the microbes.
Of course, one needs experimental tests to see if this proposed technique
works.  We are not aware of any previous studies (theoretical
or experimental) of probing the geometry of brushes in the presence of
microbes.  So we are unable to compare our numerical simulation results 
with any experimental data. 

Many issues remain to be probed. For example,
distinction of a microbe from a random colloid. We expect that
a random colloid will lead to some randomness also
in the geometry of brushes, so that the case of a cylindrical microbe
or a spherical microbe can be distinguished. For example, the reflection
symmetry of the brushes for both these cases should not be seen for a
randomly shaped particle (even angles between brushes will not be
expected to be $\pi/2$. Of course, there will not be any way to distinguish
a microbe from a dust particle of exactly same shape until one knows of
any possible differences in the anchoring angle between the two.
Also note that there is a possibility of development of a hedgehog defect 
or a Saturn ring near the central object (depending on the size of the 
central object \cite{saturn}), in that case the brush geometries
discussed here will not be seen. For very small microbe size, the 
continuum description will break down. Supposedly before that happens, 
value of scalar order parameter will also significantly change from the 
equilibrium value near the surface. We do not see that happening for the 
sizes considered here. 
 
 Our technique has not been effective in probing the size of spherical bodies.
Even with external electric field, the resulting bulge in the bands of
folded brushes has non-trivial structure too close to the spherical body.
It is not clear if it may be visible with optical microscope for the size
of body about 50 nm, like the corona virus. It is thus tempting to speculate
if the spike proteins on the surface of the corona virus can have any
non-trivial effects on the brush structure. With the size of the spike proteins 
being in ten nm range, it may be unlikely. However, the number density
of spike protein being large may possibly induce non-trivial effects
in the anchoring of nematic director at the surface. For example, if the 
director attempts to anchor around the spike protein, it will most likely
lead to very non-trivial brushes emanating out, with several boojum like
structures forming at the surface \cite{boojums}. It is not clear what type
of brush structure can emerge from that. It will be very useful to directly 
carry out experiments and see if a large distance observation involving
brushes can quickly and easily indicate presence of such tiny microbes.

\section{Acknowledgement}

I would like to thank Shikha Varma for useful discussions about microbe 
structures and possibility of testing these using nanodots and nanorods.
I also thank Sanatan Digal for useful discussions.


\end{document}